\input harvmac.tex \input tables.tex \input epsf 
\newcount\figno \figno=0
\def\fig#1#2#3{
\par\begingroup\parindent=0pt\leftskip=1cm\rightskip=1cm\parindent=0pt
\baselineskip=11pt \global\advance\figno by 1 \midinsert \epsfxsize=#3
\centerline{\epsfbox{#2}} \vskip 12pt {\bf Figure \the\figno:}  #1\par
\endinsert\endgroup\par } \def\figlabel#1{\xdef#1{\the\figno}}
\def\encadremath#1{\vbox{\hrule\hbox{\vrule\kern8pt\vbox{\kern8pt
\hbox{$\displaystyle #1$}\kern8pt} \kern8pt\vrule}\hrule}} \batchmode
\font\bbbfont=msbm10 \errorstopmode \newif\ifamsf\amsftrue
\ifx\bbbfont\nullfont
\amsffalse \fi \ifamsf \def\IR{\hbox{\bbbfont R}} \def\IZ{\hbox{\bbbfont
Z}}
\def\IF{\hbox{\bbbfont F}} \def\IP{\hbox{\bbbfont P}} \else
\def\IR{\relax{\rm
I\kern-.18em R}} \def\IZ{\relax\ifmmode\hbox{Z\kern-.4em
Z}\else{Z\kern-.4em
Z}\fi} \def\IF{\relax{\rm I\kern-.18em F}} \def\IP{\relax{\rm I\kern-.18em
P}}
\fi

  \def\tilde{\widetilde}
   
 \font\zfont = cmss10 
\def\ZZ{\hbox{\zfont Z\kern-.4emZ}} 

 %

 

\def\journal#1&#2(#3){\unskip, \sl #1\ \bf #2 \rm(19#3) }
\def\andjournal#1&#2(#3){\sl #1~\bf #2 \rm (19#3) }

\def\frac#1#2{{#1\over#2}}

\def\inbar{\,\vrule height1.5ex width.4pt depth0pt}
\def\IC{\relax\hbox{$\inbar\kern-.3em{\rm C}$}} \def\IR{\relax{\rm
I\kern-.18em
R}} \def\IP{\relax{\rm I\kern-.18em P}}    

\def\slash#1{\mathord{\mathpalette\c@ncel{#1}}} \overfullrule=0pt

\def\ZZ{{\cal Z}}  

\def\underrel#1\over#2{\mathrel{\mathop{\kern\z@#1}\limits_{#2}}}

 \catcode`\@=12



\def\nsp{{NS$^\prime$}}

\def\nsp{NS$'$}

\def\drawbox#1#2{\hrule height#2pt 
        \hbox{\vrule width#2pt height#1pt \kern#1pt 
              \vrule width#2pt}
              \hrule height#2pt}

\def\Fund#1#2{\vcenter{\vbox{\drawbox{#1}{#2}}}}
\def\Asym#1#2{\vcenter{\vbox{\drawbox{#1}{#2}
              \kern-#2pt       
              \drawbox{#1}{#2}}}}
 
\def\fund{\Fund{6.5}{0.4}}

\Title{\vbox{\rightline{hep-th/9801134} \rightline{CERN-TH/97-377}
\rightline{IASSNS-HEP-98/6}
}}
{\vbox{\centerline{On the Realization of Chiral Four-dimensional}
\centerline{Gauge Theories Using Branes}}}


\centerline{Amihay Hanany} \smallskip{\it
\centerline{School of Natural Sciences} \centerline{Institute for Advanced
Studies} \centerline{Princeton, NJ 08540, USA}} \centerline{\tt
hanany@ias.edu}

\vskip .3cm
\centerline{Alberto Zaffaroni} \smallskip{\it
\centerline{Theory Division, CERN} \centerline{CH-1211 Geneve 23, Switzerland}}
\centerline{\tt
Alberto.Zaffaroni@cern.ch}

\vskip .1in


\noindent
We consider a general brane construction for realizing chiral four-dimensional
gauge theories. The advantage of the construction is the simplicity and the possibility of realizing a large class of models existing in the literature.
We start the study of these models by determining the matter content and the
superpotential which naturally arise in the brane construction.  
\vskip 3truecm
\noindent
CERN-TH/97-377
\Date{December 97}

\lref\am{ P. S. Aspinwall and  D. R. Morrison, {\it Point-like Instantons on K3
Orbifolds},  hep-th/9705104.}
\lref\eva{S. Kachru and  E. Silverstein, {\it Chirality Changing Phase Transitions in 4d String Vacua}, Nucl. Phys.  B504 (1997) 272, hep-th/9704185.}
\lref\wpbl{M. Berkooz, R. G. Leigh, J. Polchinski, J. H. Schwarz, N.
       Seiberg and E. Witten, {\it Anomalies, Dualities, and Topology of D=6
N=1 Superstring Vacua},  Nucl. Phys.  B475 (1996) 115, hep-th/9605184.}

\lref\hm{ M. R. Douglas and  G. Moore, {\it D-branes, Quivers, and ALE
Instantons}, hep-th/9603167.}

\lref\vo{ H. Ooguri and C. Vafa, {\it Two-Dimensional Black Hole and
Singularities of CY Manifolds}, Nucl. Phys.  B463 (1996) 55, hep-th/9511164.}

\lref\gj{E. G. Gimon and C. V. Johnson, {\it  K3 Orientifolds},  Nucl. Phys.
B477 (1996) 715, hep-th/9604129.}

\lref\branes{ J. Polchinski, S. Chaudhuri and  C. V. Johnson, {\it Notes on
D-Branes},  hep-th/9602052.}

\lref\horava{ P. Horava and  E. Witten, {\it Heterotic and Type I String
Dynamics from Eleven Dimensions},  Nucl. Phys.  B460 (1996) 506,
hep-th/9510209.}

\lref\pol{ J. Polchinski, {\it Tensors from K3 Orientifolds},
Phys. Rev.  D55 (1997) 6423, hep-th/9606165.}

\lref\seibfive{N. Seiberg, {\it  Five Dimensional SUSY Field Theories,
Non-trivial Fixed Points and String Dynamics}, Phys. Lett.  B388 (1996) 753,
hep-th/9608111.}

\lref\kut{S.  Elitzur, A.  Giveon and D.  Kutasov, {\it Branes and $N=1$
Duality in String Theory}, hep-th/9702014.}


\lref\tatar{R. Tatar {Dualities in 4D Theories with Product Gauge Groups
from
Brane Configurations}, hep-th/9704198.}

\lref\kuttt{S.  Elitzur, A.  Giveon, D.  Kutasov, E.  Rabinovici and A.
Schwimmer, {\it Brane Dynamics and N=1 Supersymmetric Gauge Theory},
hep-th/9704104.}

\lref\jon{N.  Evans, C.  V.  Johnson and A.  D.  Shapere, {\it
Orientifolds,
Branes, and Duality of 4D Gauge Theories}, hep-th/9703210.}

\lref\hw{A.  Hanany and E.  Witten, {\it Type IIB Superstrings, BPS
Monopoles,
And Three-Dimensional Gauge Dynamics}, IASSNS-HEP-96/121, hep-th/9611230.}

\lref\gp{E.  G.  Gimon and J.  Polchinski, {\it Consistency Conditions for
Orientifolds and D-Manifolds}, Phys. Rev.   D54 (1996) 1667, hep-th/9601038.}

\lref\dbl{M.  Berkooz, M.  R.  Douglas, R.  G.  Leigh, {\it Branes
Intersecting
at Angles}, hep-th/9606139, Nucl. Phys.   B480 (1996) 265-278.}

\lref\kleb{ U.  H.  Danielsson, G.  Ferretti and I.  R.  Klebanov, {\it
Creation
of Fundamental Strings by Crossing D-branes}, hep-th/9705084;

O.  Bergman, M. R.  Gaberdiel and G.  Lifschytz, {\it Branes, Orientifolds
and
the Creation of Elementary Strings}, hep-th/9705130.}

\lref\barb{J.  L.  F.  Barbon, {\it
Rotated Branes and $N=1$ Duality}, CERN-TH/97-38, hep-th/9703051.}

\lref\witten{E.  Witten,
{\it Solutions Of Four-Dimensional Field Theories Via M Theory},
hep-th/9703166.}
\lref\lowe{K. Landsteiner, E. Lopez and D. A. Lowe,
{\it Duality of Chiral N=1 Supersymmetric Gauge Theories via Branes},
hep-th/9801002.} 

\lref\bh{J.H.  Brodie and A.  Hanany, {\it Type IIA Superstrings,
Chiral Symmetry, and $N=1$ 4D Gauge Theory Duality}, hep-th/9704043}

\lref\dug{M.  R.  Douglas, {\it Branes within Branes}, hep-th/9512077}

\lref\ahup{O.  Aharony and A.  Hanany, unpublished.}

\lref\ah{O.~Aharony and A.~Hanany, {\it Branes, Superpotentials and
Superconformal Fixed Points}, hep-th/9704170.}

\lref\polc{J.  Polchinski, {\it Dirichlet-Branes and Ramond-Ramond
Charges},
Phys.  Rev.  Lett.  75 (1995) 4724}

\lref\alwis{S.  P.  de Alwis, {\it Coupling of branes and normalization of
effective actions in string/M-theory}, hep-th/9705139}

\lref\witt{E.  Witten, {\it String Theory Dynamics In Various Dimensions},
Nucl. Phys.   B443 (1995) 85, hep-th/9503124.}

\lref\polstro{J.  Polchinski and A.  Strominger, {\it New Vacua for Type II
String Theory}, Phys. Lett.   B388 (1996) 736.}

\lref\karch{I.  Brunner and A.  Karch, {\it Branes and Six-Dimensional
Fixed
Points}, hep-th/9705022.}

\lref\barak{B.  Kol, {\it 5d Field Theories and M Theory}, hep-th/9705031.}

\lref\lll{K. Landsteiner, E. Lopez, D. A. Lowe, {\it $N=2$ Supersymmetric
Gauge
Theories, Branes and Orientifolds}, hep-th/9705199.}

\lref\tel{A. Brandhuber, J. Sonnenschein, S. Theisen, S. Yankielowicz,
{\it M Theory And Seiberg-Witten Curves: Orthogonal and Symplectic Groups},
hep-th/9705232.}

\lref\oog{J. de Boer, K. Hori, H. Ooguri, Y. Oz, Z. Yin,
{\it Mirror Symmetry in Three-Dimensional Gauge Theories, SL(2,Z) and
D-Brane
Moduli Spaces}, hep-th/9612131.}

\lref\LLnew{K. Landsteiner and E. Lopez, {\it New Curves from Branes}, hep-th/9708118.}
\lref\costas{C. Bachas, M. R. Douglas, M. B. Green, {\it Anomalous Creation
of
Branes}, hep-th/9705074.}

\lref\ted{A. Brandhuber, J. Sonnenschein, S. Theisen, S. Yankielowicz,
{\it Brane Configurations and 4D Field Theory Dualities}, hep-th/9704044.}

\lref\vafa{S. Katz, C. Vafa {\it Geometric Engineering of N=1 Quantum Field
Theories},  hep-th/9611090.}

\lref\vafadue{M. Bershadsky, A. Johansen, T. Pantev, V. Sadov, C. Vafa {\it
F-theory, Geometric Engineering and N=1 Dualities}, hep-th/9612052.}

\lref\vafatre{C. Vafa, B. Zwiebach, {\it N=1 Dualities of SO and USp Gauge
Theories and T-Duality of String Theory}, hep-th/9701015.}

\lref\vafaquattro{H. Ooguri, C. Vafa {\it Geometry of N=1 Dualities in Four
Dimensions} hep-th/9702180.}

\lref\ahn{C. Ahn, K. Oh {\it Geometry, D-Branes and N=1 Duality in Four
Dimensions I}, hep-th/9704061.}

\lref\ahndue{C. Ahn {\it Geometry, D-Branes and N=1 Duality in Four
Dimensions
II}, hep-th/9705004.}

\lref\ahntre{C.Ahn, R. Tatar, {\it Geometry, D-branes and N=1 Duality in
Four
Dimensions with Product Gauge Group}, hep-th/9705106.}

\lref\tatar{R. Tatar, {\it Dualities in 4D Theories with Product Gauge
Groups
from Brane Configurations}, hep-th/9704198.}

\lref\bhoy{J. de Boer, K. Hori, Y. Oz, Z. Yin,
{\it Branes and Mirror Symmetry in $N=2$ Supersymmetric Gauge Theories in
Three Dimensions}, hep-th/9702154.}

\lref\hztwo{A. Hanany and A. Zaffaroni,
{\it Branes and Six-Dimensional Supersymmetric Theories},
hep-th/9712145.}
     
\lref\popptwo{J. Lykken, E. Poppitz and S.P. Trivedi,
{\it M(ore) on Chiral Gauge Theories from D-Branes},
hep-th/9712193.}

\lref\berk{M. Berkooz,
{\it The Dual of Supersymmetric SU(2k) with an Antisymmetric Tensor and
Composite Dualities},
hep-th/9505067,
Nucl. Phys. B452 (1995) 513.}

\lref\seib{N. Seiberg, {\it Non-trivial Fixed Points of The Renormalization
Group in Six Dimensions}, hep-th/9609161; {\it Phys.  Lett.}  {\bf B390}
(1996)
753.}

\lref\dani{U.  Danielsson, G Ferretti, J.  Kalkkinen and P.  Stjernberg,
{\it Notes on Supersymmetric Gauge Theories in Five and Six Dimensions},
hep-th/9703098.}

\lref\ken{K. Intriligator, {\it RG Fixed Points in Six
Dimensions via Branes at Orbifold Singularities},  Nucl. Phys.  B496 (1997)
177-190, hep-th/9702038.}

\lref\kenju{J. D. Blum, K. Intriligator, {\it Consistency Conditions for
Branes
at Orbifold Singularities},  hep-th/9705030.}

\lref\kenjutwo{
J. D. Blum and K. Intriligator, {\it New Phases of String Theory and 6d RG
Fixed
Points via Branes at Orbifold Singularities}, hep-th/9705044.}

\lref\vaber{M. Bershadsky and C.  Vafa, {\it  Global Anomalies and Geometric
Engineering of Critical Theories in Six Dimensions}, hep-th/9703167}  

\lref\GH{O. J. Ganor, A. Hanany, {\it Small $E_8$ Instantons and Tensionless
Non-critical Strings}, hep-th/9602120, Nucl. Phys.  B474 (1996) 122.}

\lref\SW{N. Seiberg, E. Witten,
{\it Comments on String Dynamics in Six Dimensions}, hep-th/9603003,
Nucl. Phys.  B471 (1996) 121.}

\lref\AH{O. Aharony and A. Hanany,
{\it Branes, Superpotentials and Superconformal Fixed Points}, hep-th/9704170,
Nucl. Phys.  B504 (1997) 239.}

\lref\wittennew{E. Witten, {\it  New ``Gauge'' Theories In Six Dimensions},
hep-th/9710065.}

\lref\seiberg{N. Seiberg, {\it Matrix Description of M-theory on $T^5$ and
$T^5/Z_2$}, hep-th/9705221.}

\lref\intr{K. Intriligator, {\it New String theories in Six Dimensions via
Branes at Orbifold Singularities}, hep-th/9708117.}

\lref\hz{A. Hanany and A. Zaffaroni, {\it Chiral Symmetry from Type IIA
Branes},
hep-th/9706047.}

\lref\popp{J. Lykken, E. Poppitz and S. P. Trivedi, {\it  Chiral Gauge Theories
from D-Branes}, hep-th/9708134.}

\lref\bk{I. Brunner and A. Karch,
{\it Branes at Orbifolds versus Hanany Witten in Six Dimensions},
hep-th/9712143.}

\lref\ahk{O. Aharony, A. Hanany and B. Kol,
{\it Webs of (p,q) 5-branes, Five Dimensional Field Theories and Grid Diagrams},
hep-th/9710116.}

\lref\berlin{I. Brunner, A. Hanany, A. Karch and D. L\"ust,
{\it Brane Dynamics and Chiral non-Chiral Transitions},
hep-th/9801017.}

\lref\jerus{S. Elitzur, A. Giveon, D. Kutasov and D. Tsabar,
{\it Branes, Orientifolds and Chiral Gauge Theories},
hep-th/9801020.}

\newsec{Introduction}

Configurations of branes have provided a useful tool for analysing
non-perturbative properties of supersymmetric gauge theories in various
dimensions. The problem of constructing general $N=1$ chiral gauge theories
in four dimensions and of studying related problems, such as dynamical
supersymmetry breaking, is still open.

In the spirit of the construction that was initiated in \hw, there are currently
three different ways
to construct chiral gauge theories. In \bh, chiral symmetry was found
at special points in the brane realization of $N=1$ supersymmetric QCD.
This led to a localization of chiral matter in space, which was  done in \hz, but produced
only non-chiral theories. In this approach, more general constructions  lead to
chiral theories \refs{\lowe,\berlin,\jerus}.
Another construction was made in \popp, using the four-dimensional version of
the theory in \hw\
sitting at an orbifold singularity.
In this way, the theory $SU(N)^k$ with gauge factors and chiral matter associated
respectively with
the nodes and the links of the extended Dynkin diagram for $A_{k-1}$ was
realized. The theory is chiral in the sense that each one of the
bi-fundamentals connecting neighbouring factors is a chiral representation of
both the
gauge groups under which it is charged. However, each $SU(N)$ factor contains
the same number of fundamental and anti-fundamental representations.
More general theories containing tensor representations for the various
groups were obtained in \popptwo\ 
by introducing orientifold planes in the picture. The theories that
can be obtained in this way are the $N=1$ relatives of the theories classified
in \hm\ (see also \refs{\jon,\intr,\kenjutwo} for generalizations).
\lref\ooguri{J. de Boer, K. Hori, H. Ooguri and  Y. Oz, {\it Branes and Dynamical Supersymmetry Breaking}, hep-th/9801060.}

In this paper we would like to report on a third approach to construct chiral
gauge theories in this spirit.
Following
an approach used in \hztwo\ (see also \bk ) to study six-dimensional theories,
we will consider
a T-dual version of these theories, which allows more flexibility in building
models. The construction of particular models with tensor representations exists
in the literature (see, for example, \refs{\popptwo,\lowe,\berlin,\jerus}.
), but since a unified picture
in which we can realize a larger class of models is still missing, we show in
this note how to realize a big number of chiral models, 
leaving for future work the more detailed analysis of their dynamical
properties.  We will discuss what kind of superpotential is naturally present in
the brane picture. We will obtain the matter content and the superpotential of
several models, which are supposed to break supersymmetry\foot{For a recent paper dealing with the issue of supersymmetry breaking in non-chiral models constructed with branes, see \ooguri.}. 

The three approaches to the construction of chiral gauge theories
are presumably related by a sequence of T- and S-dualities. 
Below, we describe a T-duality between the second and the third.
It will be interesting to check the precise relation and to learn
more on these constructions by following the duality transformations.

We will use a mechanism proposed in \popp, which is quite general. Start with a
D-brane
realization of an $N=2$ (minimal) supersymmetric gauge theory in five dimensions.
We can obtain such a
five-dimensional theory in several ways, using, for example, D4-branes at
orbifold singularities \hm\ in Type IIA, or webs of $(p,q)$ five-branes in
Type IIB \ah.
The introduction of NS-branes, which limit the world-volume of the D-branes in
the fifth direction, following
the proposal in \hw, induces a KK reduction to four dimensions, projects out the
fields that correspond to motion in the directions transverse to the NS-brane
and generally further breaks $N=2$ to $N=1$. For the Type IIA picture
(D4-branes at orbifold singularities), the hypermultiplet matter content
of the $N=2$ theory usually parametrizes the fluctuations of the D-branes
in four spacetime directions. If the world-volume of the NS-branes is
carefully
chosen so as to freeze two of these four directions, the $N=2$ hypermultiplet
is projected out to an $N=1$ chiral multiplet. We will consider the Type IIB
realization in this paper.

There are several papers \refs{\popptwo,\lowe,\berlin,\jerus}
that deal with the brane construction for chiral theories. Some of 
the models in these papers can be connected to our construction by an explicit
T-duality.
One of the advantages of the realization presented in this paper is its
simplicity and the possibility of obtaining a very large class of models using
a unified construction.
Classical flat directions are easily studied in this approach, as in \hw, and
may be helpful in cases where the field theory analysis gets complicated.

\newsec{Building blocks for chiral theories}

A convenient way to realize an $N=1$ chiral gauge theory is to start with a
five-dimensional minimally supersymmetric $N=2$ theory, constructed with a web of
five-branes in Type IIB
\ah, and to project down to $N=1$ in four dimensions by introducing two extra
five-branes that act as boundaries in the fifth direction
\foot{Such configurations were first considered in \ahup.}.
The previous
statement is quite unrigorous. If we try to apply the set-up of \hw\ in five
dimensions, we immediately realize  that an NS-brane, having the same number of
dimensions as a D5-brane, cannot behave as a rigid boundary that would absorb the RR
charge of the D5-branes. When a D5 and an NS-brane touch \ah, they merge in a
$(1,1)$ brane, which extends in a direction dictated by supersymmetry.
Therefore, the general system is realized with $(p,q)$ five-branes that
intersect in such a way as to preserve charge and with angles dictated by
supersymmetry \ah. We can determine the branes that, in the spirit of \hw, act
as boundaries by looking at the behaviour of the system at spatial infinity. The
introduction of two extra five-branes to break supersymmetry down to $N=1$
further complicates  the construction. To simplify the description of the model
and the determination of the gauge and matter content, we will work
in this section with zero string coupling, $g_s=0$ and zero axion $\chi=0$.
This implies that we can consider the NS-branes as
infinitely rigid boundaries, which absorb the charge of the D5-branes without 
being bent. The difference in tension of the two types of branes can justify
this assumption.
To see this in more detail we consider the asymptotic orientation of a $(p,q)$
five-brane, which is restricted by the condition of supersymmetry.
Suppose that the D5-brane is stretched along a direction $x$ and the NS-brane is stretched along a direction $y$. The D5-brane is point-like in $y$ and
the NS-brane in $x$. All other directions are shared by the two
branes.
Then, a $(p,q)$ five-brane preserves the same amount of supersymmetry,
provided it is stretched as a line in the $(x,y)$ plane with a slope given by
\ahk
\eqn\slope{\Delta x : \Delta y = p + q \tau,}
where $\tau={i\over g_s}+{\chi\over2\pi}.$
In a background in which the Type IIB axion and the string coupling are zero,
any $(p,q)$ five-brane with $q\not =0$ will be parallel to the $y$ direction,
whereas any $(p,0)$
five-brane will be parallel to the $x$ direction.
This gives a support for the assumption that the NS-branes can be considered
rigid and do not bend when D5-branes end on them.
This assumption simplifies greatly the discussion about the classical field
theory with its matter content and classical interactions. At a later stage we
would like
to take the string coupling to be non-zero and then deduce information on the
quantum dynamical properties of the gauge theory studied.
But, even within this classical approximation, we will be able to make some
statements on the IR properties of the gauge theories we constructed.

In conclusion,
in this section we consider the naive T-dual of the model in \hw\ in the
presence of two extra NS-branes (which we will call \nsp). The ingredients are:
D5-branes with world-volume $(012346)$, NS-branes with world-volume $(012345)$,
\nsp\ branes with world-volume $(012367)$, and D7-branes
and O7 orientifold planes  with world-volume $(01234789)$. The two types
of NS-branes bound the D5-branes in both directions 4 and 6, and the KK
reduction in these two directions gives a four-dimensional theory. The D5
world-volume in 46 appears, in the approximation we are considering, as a
rectangle bounded by the NS-branes.
In addition to the above branes, we may introduce other branes or singularities,
which will not break  the supersymmetry further.
These are ALE space along $(4567)$, D7$'$ with world-volume $(01235689)$ and
D5$'$ with world-volume $(012357)$ \ahup. We will not make much use of these
additional branes in this paper.

The presence of these branes breaks space-time Lorentz symmetry from $SO(1,9)$
to $SO(1,3)\times SO(2)$. The first group is identified with the Lorentz
symmetry of the four-dimensional theory studied, while the $SO(2)$ symmetry acts
on the 89 directions and is identified with the $U(1)_R$ symmetry of the
$N=1$ supersymmetric system.

\fig{Brane realization of $SU(n)$ gauge theory with $n_l$ chiral multiplets in
the fundamental representation and $n_r$ chiral multiplets in the
anti-fundamental representation. The horizontal lines represent NS$'$ branes and
the vertical lines represent NS-branes. between them there are $n$
D5-branes bounded by the box. There are also semi-infinite D5-branes to the
left and to the right.}
{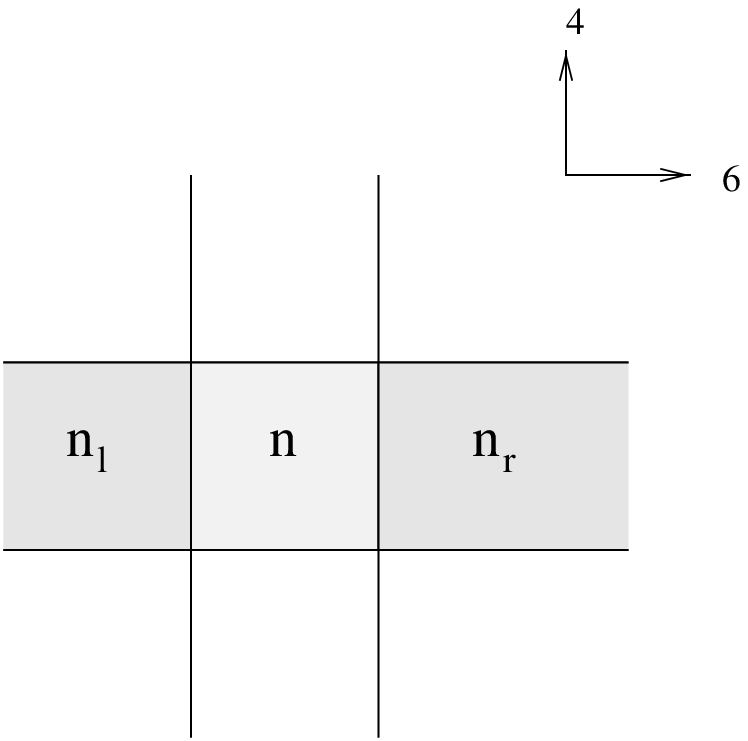}{8 truecm}
\figlabel\basic

The basic building block is obtained in the following way. Start, as in fig.
\basic, with a theory with 8 supercharges
obtained by stretching $n$ D5-branes along the direction $x_6$ between
two NS-branes and putting also $n_l$ semi-infinite D5-branes on the left and
$n_r$ on the right.
This is the five-dimensional minimally supersymmetric theory of an $SU(n)$ gauge
group with $n_l+n_r$ hypermultiplets.
Introduce further two NS$'$-branes, which bound the D5-branes at a finite distance
in $x_4$. The hypermultiplets parametrize fluctuations of the D5-branes
in 6789, but the presence of the NS$'$ freezes the possible motions in 67.
The $N=2$ hypermultiplets are projected down to $N=1$ chiral
multiplets, with different chirality depending on their position to the
left or to the right. The field content is therefore $SU(n)$ with $n_l$
fundamentals and $n_r$ antifundamentals.
Anomaly cancellation requires that $n_l=n_r=n_f$. It would be interesting to see
how this arises from RR charge conservation.

The parameters of the theory are given by the positions of the two types of
NS-branes. There are 4 NS-branes with 4 transverse positions to each, which
gives 16 possible parameters. Six parameters can be set to zero by tuning the
origin in directions 456789. We are left with 10 parameters. There are 4
distances in the 4567 directions and there are 3 positions in the 89 directions.
The gauge coupling is given by the area of the rectangle in the 46 directions:
\eqn\gaco{{1\over g^2}={\Delta x_4 \Delta x_6\over g_s l_s^2}.}
The other parameters do not have a clear interpretation. The distances in the
5 and 7 directions look like FI terms for the $U(1)$ gauge groups. Since it is
frozen from a four-dimensional point of view, they may be promoted to moduli of
the field theory. The same comment applies for the other positions in the 
directions 89. An analysis of specific cases makes their interpretation clearer,
mostly as dynamical moduli.

The theta angle is related to the Type IIB axion.
The D5-brane has a term in the effective action, which looks like
\eqn\sixtheta{{\chi\over2\pi}F\wedge F\wedge F.}
Integrating over the rectangle in the 46 directions, the four-dimensional theta
angle is given by
\eqn\fourtheta{{\theta\over2\pi}={\chi\over2\pi}\int_{46}F.}

Let us study some classical flat directions.
Let us denote the fundamental fields by $Q_i$ and the antifundamental fields by
$\tilde Q^j$.
We can reconnect a left semi-infinite D5-brane to a finite D5-brane and to a
right semi-infinite D5-brane. They form an infinite D5-brane in the 6 direction.
The D5-brane is now free to move between the two NS$'$ branes in the 7
direction. What remains are $n-1$ finite branes with $n_f-1$ semi-infinite
branes to the left and to the right. The gauge group is broken to $SU(n-1)$ with
$n_f-1$ flavours. Such a motion corresponds to a non-zero expectation value for a
meson field, say $\tilde Q^1Q_1$. One can generalize this to $r$ such branes.
The gauge group is broken to $SU(n-r)$ with $n_f-r$ flavours. The positions of
the branes parametrize the eigenvalues of the mesonic matrix
$M_i^j=\tilde Q^jQ_i$. For this case, $r$ of them are non-zero.
For $n_f\ge n$, there is also a baryonic branch. This corresponds to
reconnecting at one side, say the right, $n$ of the semi-infinite D5-branes.
The $n_f-n$ remaining semi-infinite branes can now move, together with the right
NS-brane, along the 7 direction. The distance in the 7 direction corresponds to
the expectation value of the baryons. This case serves as an example to the
comment made in the last paragraph on distances between NS-branes.

One may question that the matter localized at the intersection of two D5-brane
along the NS-branes bounded by NS$'$-branes is indeed chiral and not hyper.
The agreement of the classical moduli space with the field theory in question
serves as a support for this identification.

\fig{Brane realization of product of $SU$ gauge groups.
The horizontal lines represent NS$'$ branes and
the vertical lines represent NS-branes. There are $n_l$ ($n_r$) semi-infinite
D5-branes to the left (right). Finite D5-branes are bounded by the NS-branes and
the NS$'$-branes. The number of D5-branes is denoted in each box. The different
shaded regions are to emphasize that the number of D5-branes can be different.}
{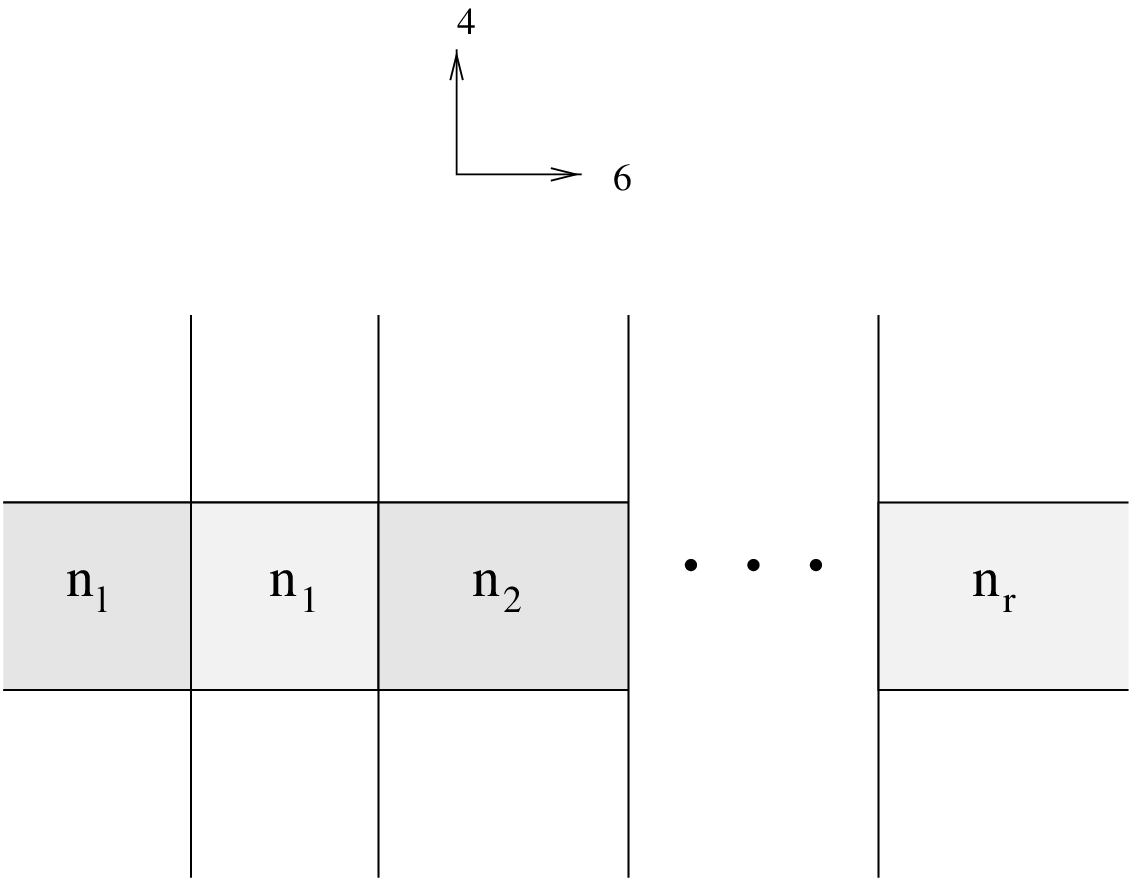}{10 truecm}
\figlabel\genbas

It is easy to generalize the theory to a product of $SU$ gauge factors.
Consider, as in fig. \genbas,
$P$ NS-branes displaced along the direction $x_6$, with $n_i$ D5-branes in
between the $i$-th and $(i+1)$-th NS-brane and $n_l$ semi-infinite D5-branes on
the left and $n_r$ on the right. The $N=1$ gauge theory is
\eqn\th{\prod_{i=1}^{P-1} SU(n_i)}
with chiral bifundamentals charged under each neighbouring factor, and extra
$n_l$ fundamentals for $SU(n_1)$ and $n_r$ antifundamentals for $SU(n_{P-1})$. 

For general values of $n_i$ the theory is obviously anomalous. Charge
conservation for the RR spacetime fields should be correspondingly violated.
Since we have discarded the bending of the five-branes and the issue of the
charge conservation at the intersection of various branes, we cannot see
 this spacetime phenomenon explicitly.  For the moment, we use the field theory
input.
There are two series of anomaly-free models. The first one has $n_l=n_i=n_r=n$,
\eqn\thone{\prod_{i=1}^{P-1} SU(n)}
with chiral bifundamentals, $n$ fundamentals for the first factor and
$n$ antifundamentals for the last. This is the T-dual of the model in \popp,
which can be exactly reproduced if we take a compact $x_6$ direction. The second
anomaly-free model  has the following field content:
\eqn\thtwo{SU(n)\times SU(m)\times SU(n)\times SU(m)\times ...}
with chiral bifundamentals and $m$ fundamentals for the first factor. The
last factor can be $SU(n)$ with $m$ antifundamentals or  $SU(m)$ with $n$
antifundamentals.

We can introduce extra matter using D7-branes. Each of them produces a
hypermultiplet in the $N=2$ theory coming from the open strings between D5- and
D7-branes. Since the intersection between D5- and D7-branes is, in general,
localized far from the D5 boundaries, the full multiplet, which decomposes in
two chiral multiplets of the $N=1$ theory, survives the projection imposed by
the NS-branes. We will give further consistency checks on this identification in
the next section by using brane motion and creation.
The D7-brane gives rise to two scalars, positions in the 56 directions.
The position in the 5th direction gives rise to a mass for the quark multiplet.
Together with the Wilson line along the  direction 4 of the D7-brane, they
combine into a chiral multiplet, which gives rise to the mass of the quark fields.

The superpotential of the model is derived from the local $N=2$ supersymmetry.
It gives a Yukawa coupling with an adjoint field and two quark fields.
However, the adjoint scalar involved is frozen by the bounding by NS-branes.
This leads to a configuration with no superpotential. A mass term for the
quarks still exists.
The superpotential can be modified by introducing a rotation of the D7-branes in
the  directions 47-56. This changes the coefficient of the Yukawa coupling as
in cases discussed in \ah.

We can construct higher-order representations by introducing an orientifold
plane parallel to the D7 in the
picture. There are two possible choices of sign for the charge of such an
orientifold plane: in the case of a negative sign, which arises in the standard
Type I$'$ string theory, we call the orientifold O7, while in the case of a
positive sign, we call it O7$^+$. 
Every D- or NS-brane now must have an image under the $Z_2$ symmetry
$x_{5,6}\rightarrow -x_{5,6}$ or be stuck at the orientifold point. We will
take the two NS$'$  stuck at $x_5=0$ and the NS-branes disposed in $x_6$
in such a way as to preserve the $Z_2$ symmetry. There are essentially two
basic configurations \refs{\LLnew,\hztwo,\bk},

\fig{$Sp(n) (SO(2n))$ gauge group with $n_f$ flavours. The dashed line
represents an O7 (O7$^+$) plane. $n_f$ semi-infinite D5-branes give rise to
$n_f$ chiral multiplets.}
{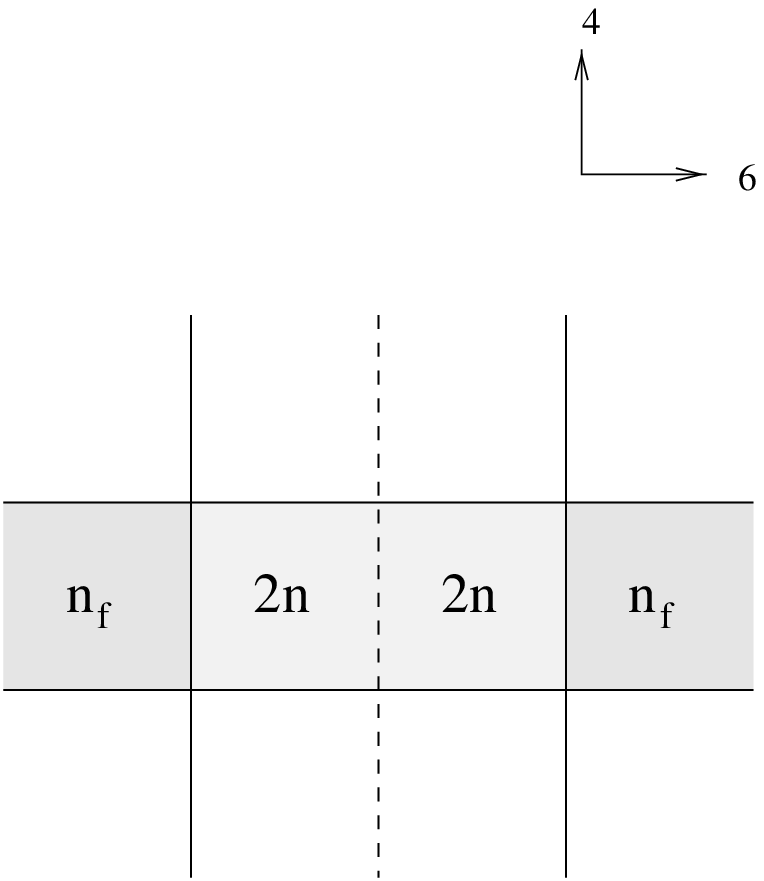}{10 truecm}
\figlabel\Sp

\item{1)} Consider, as in fig. \Sp, an NS-brane at $x_6^0$ and stretch $2n$
D5-branes between
it and its image at $-x_6^0$. Put also $n_f$ semi-infinite D5-branes on the
right of the NS-brane. The two NS$'$-branes project down the theory to
$USp(2n)$ (for O7) or $SO(2n)$ (for O7$^+$) with $n_f$ fundamentals. In the O7$^+$
case, there is the option to have an odd number of D5-branes, one of them being
stuck, giving the group $SO(2n+1)$.
\item{2)} Consider an NS-brane stuck at $x_6=0$ and stretch $n$ D5-branes in
between it and a second NS-brane at $x_6^0$. Put also $n_f$ semi-infinite
D5-branes on the right
of the second NS-brane. Since the D5-branes are identified with their images
(D5-branes stretched between the stuck NS-brane and the image
of the second NS at $-x_6^0$), no projection on the Chan-Paton factors is needed
and the gauge group is $SU(n)$. The open strings connecting the D5-branes with
their images give rise to a multiplet in the antisymmetric
representation of the gauge group (for O7) or in the symmetric one (for O7$^+$)\LLnew.
In conclusion, the theory is $SU(n)$ with an antisymmetric (symmetric) and
$n_f$ antifundamentals.

The theory in 1) is well defined for all values of $n$.
Global anomalies of $Sp$ gauge theories give rise to a constraint on $n_f$.
In the case of $O7$, $n_f$ must be even.
For the case of $O7^+$, the gauge group is $SO$ and there is no restriction on
$n_f$.
For the theory 2), anomaly cancellation requires $n=n_f+4$ (for O7) and
$n=n_f-4$ (for O7$^+$).

We can now summarize two rules for D5-branes stretched between a pair of
NS-branes and a pair of NS$'$-branes. The rules are derived from the field
theory requirement that the anomalies will be cancelled. It would be nice to get
this rule from an independent argument, which does not rely on the field theory
analysis. Nevertheless, we shall state the rules and study the consequences of
these two rules below. We hope to return to an independent derivation of these
rules in the future.
The rules are:
\item {a)} Given, as in fig. \basic, $n$ D5-branes between two NS-branes
and two NS$'$-branes with $n_l(n_r)$ D5-branes to the
left (right), the consistency condition requires that $n_l=n_r$.
\item {b)} Given, as in fig. \Sp, $n$ D5-branes between NS-brane and its
image under an O7 (O7$^+$) and between two NS$'$-branes with $n_f$ D5-branes
connecting along the NS-branes, the consistency condition requires that
$n_f=n-4(n+4)$.
We will use these rules below.

Generalizations can be obtained by adding other NS-branes (and their images
under $Z_2$). If $n$ D5-branes are stretched between two NS-branes that are
not stuck at the orientifold, the $Z_2$ projection identifies them with their
images living on the other side of the orientifold, without further projecting
the Chan-Paton factors. Therefore they give rise to $SU(n)$ gauge groups. The
generalization of the theories in 1) and 2) (for O7) reads, respectively,
\eqn\vectwo{\eqalign{USp(V_0)&\times SU(V_1)\times SU(V_2)\times ... ,\cr
SU(V_0)&\times SU(V_1)\times ...}}
with chiral bifundamentals for each neighbouring gauge group, an antisymmetric
for the factor $SU(V_0)$ and $n_f$ antifundamentals for the last factor if
we put $n_f$ semi-infinite D5-branes. 
In all the above cases, the groups $USp$ become $SO$ if we use
O7$^+$.

We conclude this section by discussing the relation of the Type IIB construction with the one considered in \refs{\popp,\popptwo} using branes at orbifold singularities. We will see that an explicit T duality maps the models in \refs{\popp,\popptwo} in a subset of the theories we have considered in this section.

The theory in \popp\ is obtained by starting with $N$ D4-branes at a $Z_k$
orbifold singularity. We take the world-volume of the D4-branes to extend in
the directions $(01234)$ and the orbifold projection to act on $(6789)$. The
world-volume theory was determined in \hm\ and is associated with the extended
Dynkin diagram for $A_{k-1}$: each of the nodes provides an $SU(N)$ gauge factor
and each of the links a hypermultiplet in the bifundamental of the two
gauge groups corresponding to the nodes connected by the particular link.
The hypermultiplets parametrize the fluctuations of the D4-brane positions
in $(6789)$. The introduction of two NS-branes with world-volume $(012367)$
and different $x_4$ positions
induces a KK reduction along the direction 4, breaks the supersymmetry down to
$N=1$ and freezes the 89 scalars in the hypermultiplets. Notice that the
NS-branes must be stuck at $x_8=x_9=0$ to respect the $Z_k$ projection without
introducing images and extra states in the theory. The surviving matter is
composed by chiral multiplets parametrizing the motion in 67. For each $SU(N)$
factor there are two such chiral multiplets corresponding to the two links
that connect the given node to the two adjacent ones. They are fields in the
fundamental representation of the gauge group, but with opposite chirality.
The theory is anomaly-free since each gauge factor has the same number of
fundamentals and antifundamentals.

Extra matter can be introduced using D8-branes with world-volume $(012346789)$,
which do not break any further supersymmetry. Each of them provides a pair
of chiral multiplets in the fundamental and antifundamental representation.
If we further introduce an O8 orientifold plane parallel to the D8-branes, the
gauge and matter content is projected out according to the rules in \hm. The
resulting theories have the form:
\eqn\vect{\eqalign{USp(V_0)&\times SU(V_1)\times SU(V_2)\times\cdots
SU(V_{P-1})
\times USp(V_P),\,\,\, k=2P,\cr USp(V_0)&\times SU(V_1)\times
SU(V_2)\times\cdots
SU(V_{P-1})\times SU(V_P),\,\,\, k=2P+1,\cr SU(V_0)&\times SU(V_1)\times\cdots
SU(V_{P-2})\times SU(V_{P-1}),\,\,\,
k=2P.}}
The two cases for $k$ even correspond to different kinds of projections.
The matter consists of chiral bifundamentals charged under each neighbouring
gauge factor and of chiral anti-symmetric for the first and last $SU$
factors\foot{The theories in \popptwo\ are obtained using an O4 plane instead of
an O8 plane. The matter content of these theories is again classified in \hm.}.
It is convenient to perform a T-duality
and transform the system to a brane configuration in Type IIB. This approach
was already used in \hztwo\ and \bk\ to study six-dimensional theories.

If we perform a T-duality in $x_6$, the $Z_k$ orbifold singularity is converted
into a set of $k$ NS-branes with world-volume $(012345)$ positioned along $x_6$.
The D4-branes becomes D5-branes that can break between the new NS-branes.
The two old NS-branes with world-volume $(012367)$ remain unchanged. These are
exactly the models we discussed before. It is easy
to check that the gauge and matter content remains the same. The theories in
\vect\ can be obtained by
using a compact $x_6$ direction. The three different theories in \vect\
correspond to one odd case and two possible ways to put $2P$ NS-branes on a
circle in a $Z_2$ invariant way \refs{\hztwo,\bk}.

The Type IIB description is more general. Not all the realizations of models
have known T duals in a description with branes at orbifold singularities.
Nevertheless, the Type IIA picture
is a complementary description and may be sometimes useful. In particular,
 the matter content and the superpotential of the Type IIA models can be
explicitly computed using the orbifold projection. This provides non-trivial
consistency checks about the Type IIB description. In the cases in which
the theories have a known T dual we can check the correctness of the proposed
matter content and superpotential by an explicit computation in the T-dual
picture.

\newsec{Arrays of D5 boxes}

\fig{More general construction of a product of $SU$ gauge groups.}
{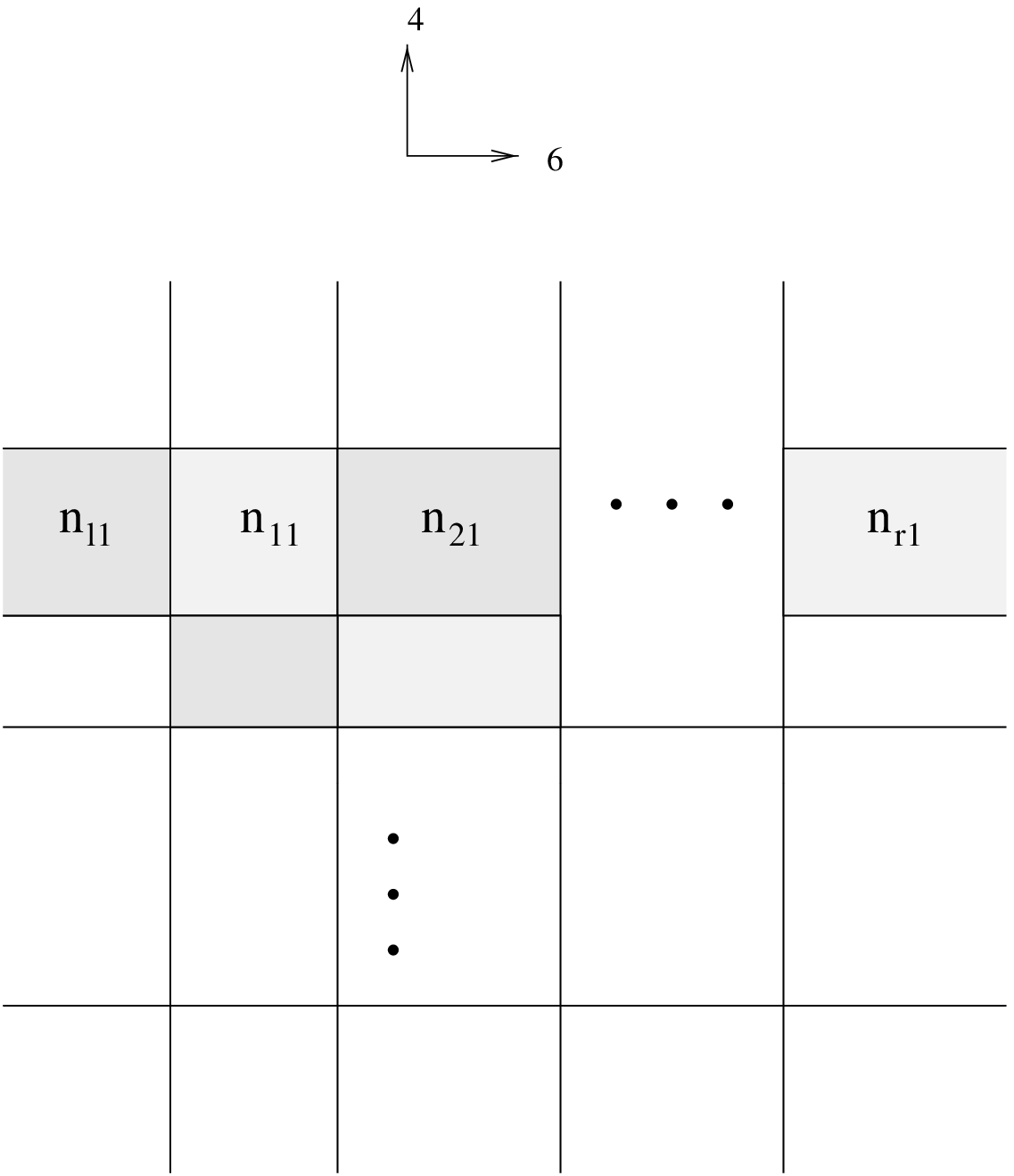}{10 truecm}
\figlabel\grid
Another generalization is to consider a grid of SU theories as in fig. \grid.
We take $P$ NS-branes and $R$ NS$'$-branes. Let the matrix $N=\{n_{ij}\}$,
$i=0,\ldots,P$, $j=0,\ldots,R$, denote the number of D5-branes in each box.
Then, the gauge group is
\eqn\grr{\prod_{i=1}^{P-1}\prod_{j=1}^{R-1}SU(n_{ij}).}
We expect chiral bifundamental matter at the intersection of any two boxes and
chiral multiplets at the boundaries. Boxes can intersect along a line (for
example, $n_{11}$ and $n_{12}$) or at a point (for example, $n_{12}$ and
$n_{21}$).
In both cases, we can expect chiral bifundamentals.
The projection imposed by
the NS-branes gives the following matter content for the model: there are, in
horizontal, chiral bi-fundamentals $(n_{i,j},\bar n_{i,j+1})$, in vertical
$(n_{i,j},\bar n_{i+1,j})$ and along the diagonal $(\bar n_{ij},n_{i+1,j+1})$.

\fig{A three-box model with three groups, three chiral multiplets and a cubic
superpotential. The number of D5-branes in each box is denoted by an integer.
Empty boxes have no bound D5-branes.}
{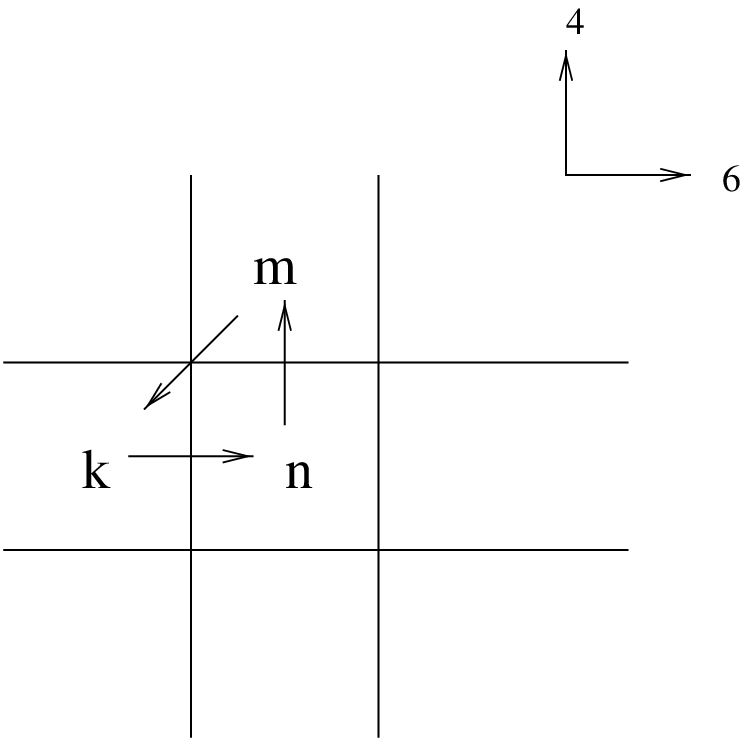}{10 truecm}
\figlabel\tbox

Let us consider the three-box model of fig. \tbox.
This model will serve as the basic building
block for the models we will consider below.
We can picture, as in fig. \tbox, the chiral bifundamentals that connect
neighbouring boxes with arrows. Their direction specifies the chirality of the
bifundamental. The rule is that the arrows can only be drawn in the direction 
East, North and South-West.
The gauge group is $SU(k)\times SU(n)\times SU(m)$. The boxes may have finite or
infinite area, which will correspond to a gauge symmetry or a global symmetry,
respectively. This property, however, does not enter into the present
discussion.
As the arrows indicate, there are three chiral multiplets $Q_1, Q_2, Q_3$,
transforming in $(k,\bar n, 1), (1,n, \bar m), (\bar k, 1, m)$, respectively.
These fields allow for a superpotential of the form
\eqn\tre{W=Q_1Q_2Q_3.}

Why is this the matter content? 
Previous experience with simple models such as those in fig. \genbas\ teaches us
about the horizontal bi-fundamentals and, using the symmetry of the
construction, about the vertical ones.
The presence of diagonal fields is more subtle. There are several arguments
for their existence and to explain why only the fields associated with one of the two
diagonals survive.
First, recall that the chiral multiplets are given by strings that are
stretched between the two D5-branes. In the absence of NS-branes
(here, we shortly refer to either an NS or an NS$'$-brane), two
possible orientations are allowed for the strings, which correspond to two
opposite chiral fields. The presence of NS-branes induces a particular 
orientation for the strings. This gives rise to a single chiral field, with a
given chirality. A string can be only parallel to the NS-brane and
not antiparallel. This is why only an orientation, say going east (the choice
of orientation is a matter of convention, and is specified by choosing the
orientation on the NS-branes), is allowed and not, say, going west. Such strings are
parallel to the NS-branes (and not to the NS$'$-branes).
Similarly, only an arrow going, say, north and not going south can produce a
multiplet. The string is then parallel to the NS$'$-branes.
The diagonal intersection is restricted by the orientation of both the NS- and
NS$'$-brane and goes only south-west.
In addition, since the three arrows form a closed circle, they give rise a
superpotential, eq. \tre.

\fig{A creation of various boxes of D5-branes when a D7 and an NS-brane cross.
The D7-brane is represented by a vertical dotted line.}
{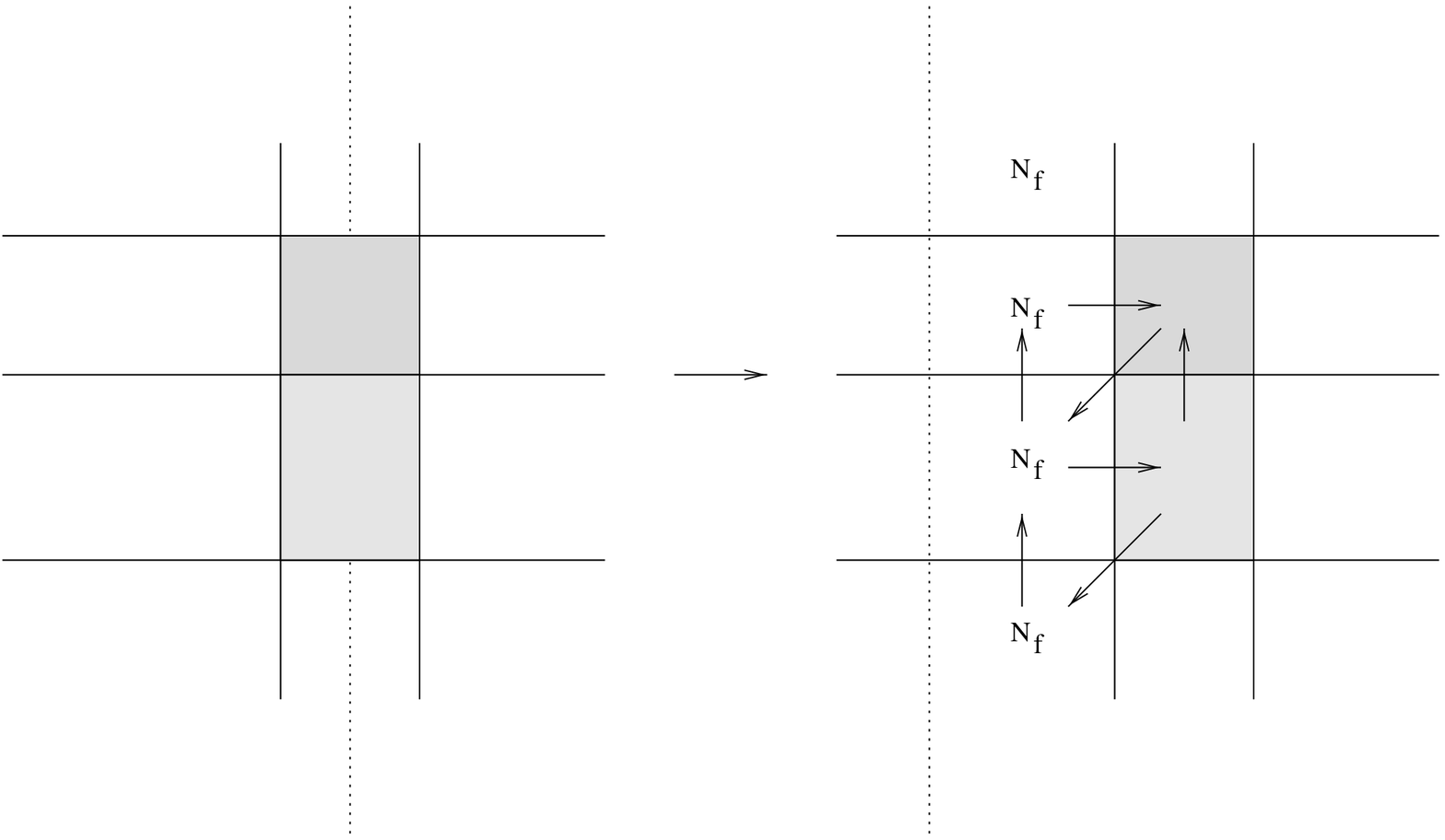}{14 truecm}
\figlabel\creation

A second argument goes as follows. Consider, as in fig. \creation,
the simple case in which we have
a column of boxes, for example, $n_{ij}=\delta_{jP}n$ for some $P$ with the
addition of a
D7-brane. The D7-brane provides vector-like matter for all the gauge groups.
We can move the D7 to the left.
When the D7 crosses the leftmost NS-branes,
other D5-branes are created between it and the NS-brane \hw.
Next move the D7 to infinity.
The matter in the fundamental is now provided by semi-infinite
D5-branes $n_{j-1,P}=1$. We shall assume, as in previous cases following \hw,
that the matter content does not change in this transition.
There are chiral fundamentals provided by the semi-infinite D5-branes.
To have a vector-like matter as before the phase transition, we
need antifundamentals coming from one and only one diagonal.
The discussion on D7-branes is expanded in the next subsection.

As a third check, we notice that, by performing a T-duality, some of the models
can be identified with the ones discussed in \popp, where the matter content
can be explicitly derived by an orientifold computation. The two results agree.

But perhaps the best motivation for the above matter content is the fact that
the flat directions expected in field theory are exactly matched by the allowed
motion of the branes. We will see several examples below. We must postpone the
discussion until we specify the superpotential for all these theories. 

\subsec{Matter and Interactions from D7-branes}

In this subsection we study in detail the introduction of D7-brane to the
system. We find that there are, in addition to the expected results,  new
interactions that appear. For this reason, we devote this subsection to this
effect.

\fig{The creation of various boxes of D5-branes when a D7-brane and an NS-brane cross.
The case with a single finite box. The arrows indicate all possible chiral
fields relevant to the interaction between the D7-brane and the D5-branes.}
{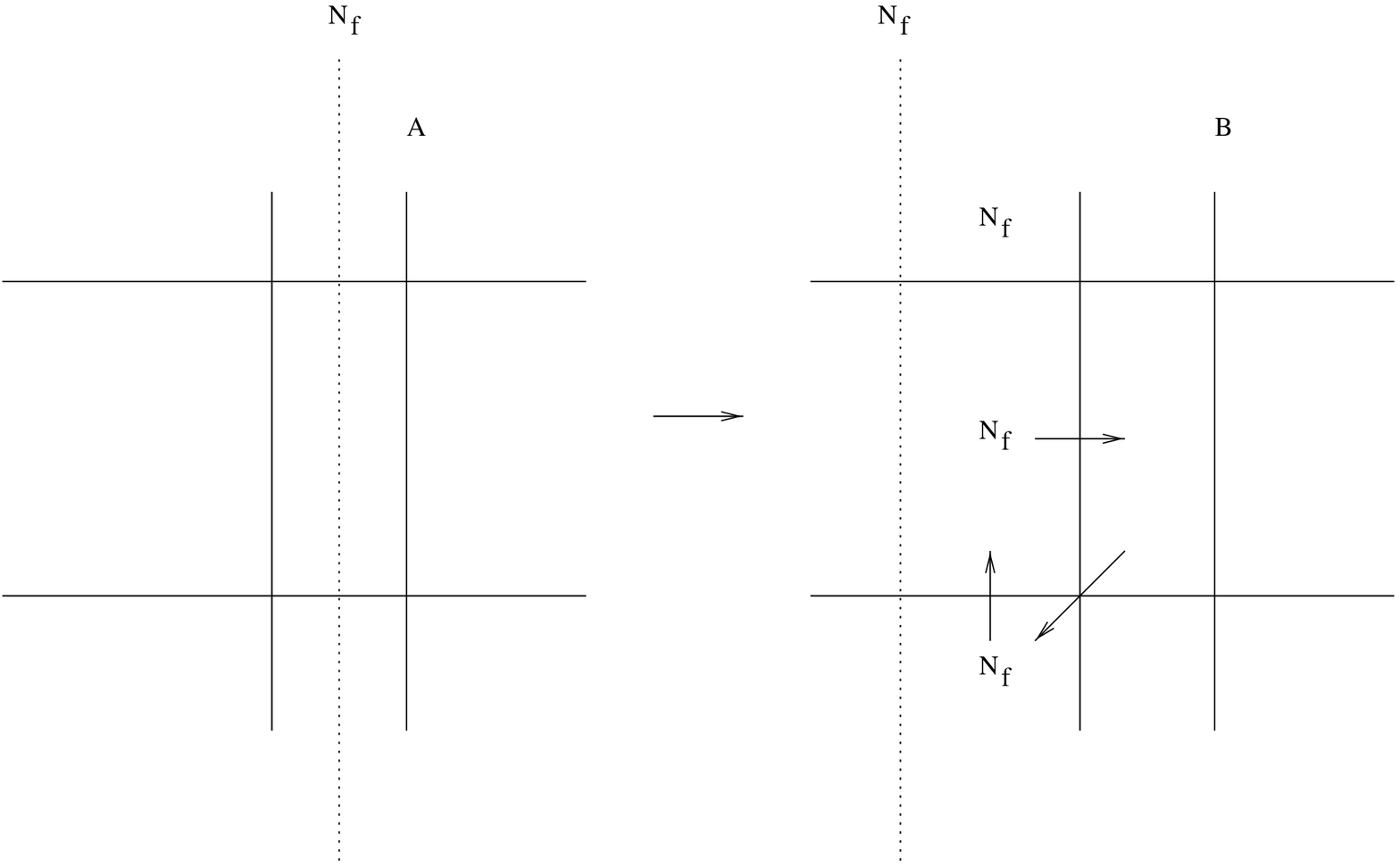}{14 truecm}
\figlabel\creone

Consider, as in fig. \creone, a single box with $n$ D5-branes and a
D7-brane. As in the usual case, the D7-brane gives rise to two chiral multiplets
$Q, \tilde Q$, with opposite chirality and a superpotential given by
\eqn\lupo{m \tilde Q Q,}
where $m$ is given by the distance of the D5- from the D7-brane in the 5
direction (together with a complex partner).
As mentioned above, we also assume that when a D7-brane crosses an NS-brane, the
matter and interactions are not changed.
Let us indeed move the D7-brane to the left. A D5-brane is stretched between
the D7 and left NS-brane. The 5 position of the D7-brane can be tuned to touch
the D5-branes. Then there are three boxes of D5-branes, as in fig.
\creone B.

We would like to look for fields and interactions as in eq. \lupo.
Using the rules of fig. \tbox\ and eq. \tre, we find that in addition to
the quark fields, $Q, \tilde Q$, denoted by the diagonal and horizontal arrows
in fig. \creone B,
respectively, there is another field $m$ that transforms in the
bifundamental of the global symmetry, denoted by the vertical arrow in fig.
\creone B. The superpotential is related to the upper closed triangle of
arrows and leads to eq. \lupo.
We find agreement with the expectations.
Let us next move the D7 to the right.
By symmetry, we should find the same matter and superpotential.
One difference is that the interaction now comes from a lower closed triangle.
We learn that any closed triangle of arrows should contribute to the
superpotential.
In conclusion, by looking at the one-box case, we find agreement with the
usual interaction familiar from theories with 8 supercharges.

Let us look at the two-box case.
We read off the matter content from fig. \creation,
which consists of the expected quark fields $Q, \tilde Q$ and $Q', \tilde Q'$
and the mass terms $m$ and $m'$. However, in addition to these fields there is
a new field $F$ that transforms in the bifundamental of the two gauge groups.
The interaction can be read off from the arrows in the figure.
We first note that there are two upper triangles and one lower triangle.
The two upper triangles give rise to the usual interactions that are present in
the one-box case.
The lower one gives a new term for a configuration of D7 with more than one box.
To summarize, there are three terms in the superpotential
\eqn\tupo{m \tilde Q Q + m' \tilde Q' Q' + F Q' \tilde Q.}
If we repeat the analysis for a motion of the D7-brane to the right, we get a
similar term, the contribution comes from two lower triangles and one upper
triangle.
As a nice check of eq. \lupo\ and \tupo, we can remove the middle NS$'$-brane away in the 45 directions. This gives a mass to the fields $Q'$ and
$\tilde Q$ in eq. \tupo, which upon integration leads to eq. \lupo.
The case with more than two boxes is generalized in an obvious way and does not
give new effects.

By assuming that the D7-brane motion in the $x^6$ direction is an irrelevant
parameter to the field theory, we conclude that the presence of a D7-brane in
addition to some boxes of D5-branes gives rise to matter fields as in the
previous paragraph and a superpotential given by eq. \tupo.

\newsec{Interesting models and their properties}

The theories we just constructed have several non-trivial IR properties, the
most interesting one being probably the fact that they can exhibit supersymmetry
breaking. We hope that the techniques we just described can be useful to better understand
the non-trivial dynamics of these theories. An immediate question is Seiberg's duality for these models. With the
explicit realization of chiral theories using D- and NS-branes, the technique in
\refs{\hw,\kut} can be immediately applied to the models. Even
more interesting would be to demonstrate that some of these models indeed break
supersymmetry. We leave the real
hard questions for future work, hoping that our technique will prove useful
to get a better understanding of the IR properties of chiral theories, and,
for the moment, we limit ourselves to some simple considerations and consistency
checks.

We also want to note that up to now we assumed that all the $U(1)$ factors
are frozen. Since they are generally anomalous, this must be so. However,
as already noticed in \popp, it may happen that, in order to get agreement
between the field theory flat directions and the allowed brane motions, we have
to impose their D-term equations. Analysis
of specific cases may help in understanding the role of the $U(1)$ factors. However, since we expect that quantum corrections to the brane configuration have a lot to say about the fate of these $U(1)$ factors, we cannot make  a general statement about them at this level. In the examples presented below,
they do not seem to play any role (see however the remark at the end of section 4.2) .

\subsec{Superpotential}

None of the previous models is completely defined until we say if there is a
superpotential and what form this has. We will give a general rule for reading the
superpotential out of the brane construction. With such a superpotential
the flat directions derived from field theory are exactly matched by the
allowed motions of the branes. In the cases in which the model has a known T-dual
description and becomes one of those in \popp, the proposal agrees with the
explicit computation.

There is a simple guideline in searching for the superpotential. We do not
expect any superpotential in the theories in fig. \genbas. The open strings giving
chiral bifundamentals are localized near the intersection of the boxes.
Being localized at different points, we do not expect any interaction between
different bifundamentals. On the other hand, in the theories of figs. \grid\
and \tbox, the open strings giving rise to the bifundamentals $(n_{ij},
\bar n_{i,j+1})$ ($Q_1$), $(n_{i,j+1}, \bar n_{i+1,j+1})$ ($Q_2$) and
$(n_{i+1,j+1},\bar n_{ij})$ ($Q_3$) can touch at the corner of three boxes, and
we can expect a 
superpotential. The general rule is that, every time the open strings can
interact, there is indeed a superpotential. The basic building block is depicted
in fig. \tbox\ and gives rise to the superpotential
\eqn\spt{W=Q_1Q_2Q_3.}
We have a superpotential for each closed cycle made up by arrows, as in
fig. \tbox.

The main reason for the existence of this superpotential is the fact that the flat directions predicted by the field theory analysis coincide with the
allowed motion of the branes. 
We see many examples that demonstrate the exact matching between flat directions and motion of branes in the next sections, when we will construct
interesting models.

As a consistency check, we can consider models for which the T-dual description in Type IIA as branes at orbifold singularities is known and the superpotential 
can be explicitly computed. In particular, the models in \popp\ can be realized 
and the results of the two methods compared. One finds complete agreement.

\lref\aff{I. Affleck, M. Dine and N. Seiberg, {\it  Dynamical Supersymmetry Breaking In Four Dimensions And Its Phenomenological Implications},  Nucl. Phys.  B256 (1985) 557.}
\lref\thin{K. Intriligator and S. Thomas {\it  Dynamical Supersymmetry Breaking on Quantum Moduli Spaces} Nucl. Phys.  B473 (1996) 121, hep-th/9603158.}
\lref\thintwo{K. Intriligator and S. Thomas {\it Dual Descriptions of Supersymmetry Breaking}, hep-th/9608046.}
\lref\mura{ H. Murayama, {\it  Studying Non-calculable Models of Dynamical Supersymmetry Breaking}  Phys. Lett.  B355 (1995) 187,  hep-th/9505082.}
\lref\dine{M. Dine, A. E. Nelson, Y. Nir and Y. Shirman,  {\it New Tools for Low Energy Dynamical Supersymmetry Breaking}  Phys. Rev.  D53 (1996) 2658, hep-ph/9507378.}
\lref\shad{E. Poppitz, Y. Shadmi and S. P. Trivedi {\it Duality and Exact Results in Product Group Theories}, Nucl. Phys.  B480 (1996) 125,  hep-th/9605113.}
\lref\pos{E. Poppitz and  S. Trivedi, {\it Some Examples of Chiral Moduli Spaces and Dynamical Supersymmetry Breaking},  Phys. Lett.  B365 (1996) 125, hep-th/9507169.}
\lref\rand{C. Csaki, L. Randall and W. Skiba, {\it More Dynamical Supersymmetry Breaking},Nucl. Phys.  B479 (1996) 65, hep-th/9605108.}
\lref\randtwo{ C. Csaki, L. Randall, W. Skiba and R. Leigh, {\it Supersymmetry Breaking Through Confining and Dual Theory Gauge Dynamics},  Phys. Lett.  B387 (1996) 791, hep-th/9607021.}
\lref\pouliot{  P. Pouliot, {\it  Duality in SUSY $SU(N)$ with an Antisymmetric Tensor}, Phys. Lett.  B367 (1996) 151,  hep-th/9510148.}
\lref\postwo{ E. Poppitz, Y. Shadmi and S. P. Trivedi {\it Supersymmetry Breaking and Duality in SU(N)xSU(N-M) Theories},  Phys. Lett.  B388 (1996) 561, hep-th/9606184.}
\lref\ratt{R. G. Leigh, L. Randall and R. Rattazzi {\it  Unity of Supersymmetry Breaking Models},  Nucl. Phys.  B501 (1997) 375, hep-ph/9704246.}

\subsec{The (3,2) model}
In this and the following sections, we construct several models that are supposed to break
supersymmetry \aff\ (see \refs{\mura,\dine,\pos,\pouliot,\thin,\rand,\shad,\postwo,\randtwo,\thintwo,\ratt} for examples and discussions in the recent literature). Almost all the {\it classical} cases can be realized with the
technique
presented in the previous sections. One important point is that models with a
particular matter content can be realized in several ways using the rules that
we gave in section 3 and 4. However, different dispositions of boxes can give
the same matter content but, generally, different superpotentials. Since we want
models that break supersymmetry, we are interested in theories without flat
directions. In the brane picture, this condition translates to the statement
that no brane can move away from the system without spoiling the equilibrium
or violating some charge conservation.

We also restrict our discussion to a classical analysis of the models. The study
of full quantum corrections requires taking non-zero string coupling. We hope to
return to this point in the future.

\fig{The (3,2) model $SU(3)\times SU(2)$ with chiral fields and a
superpotential.}
{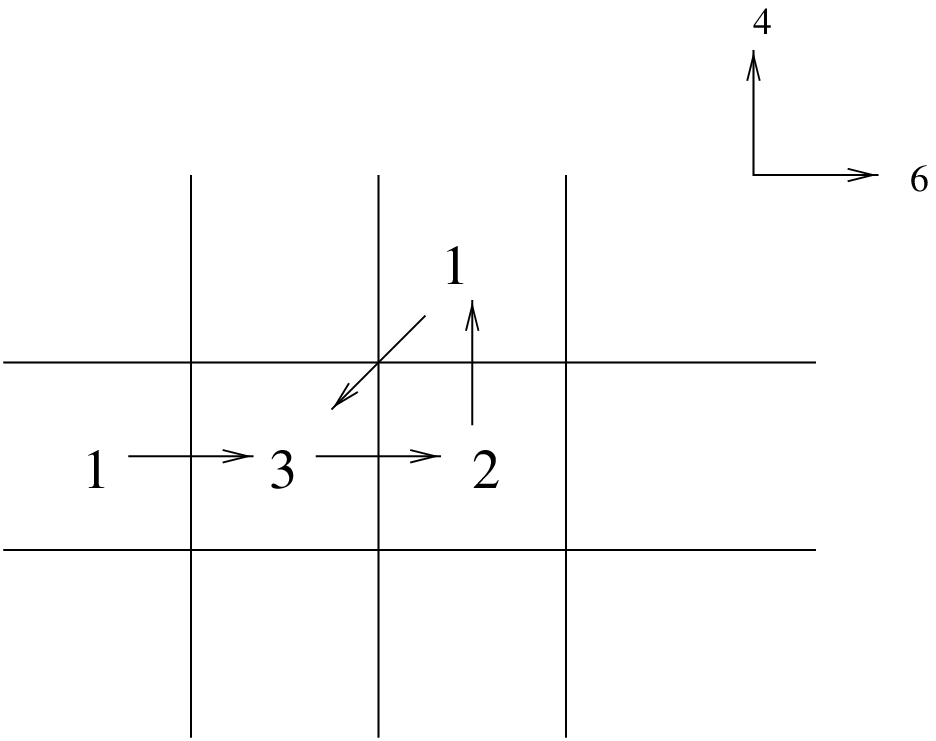}{8 truecm}
\figlabel\tt

Consider the configuration in fig. \tt. The gauge group is
$SU(3)\times SU(2)$. The chiral fields are $R_i$, $i=1,2$, transforming in
$(\bar3,1)$, $L$ transforms in (1,2) and $Q$ in (3,2).
The model is anomaly-free.
Using the three-box model in fig. \tbox, there is a superpotential
\eqn\thtw{W=R_1QL.}
The model is supposed to break supersymmetry, which will be seen by taking
non-zero string coupling $g_s$. The model has no flat directions.
It is immediate to see that indeed there is no allowed motion for the branes in
fig. \tt.

\fig{The (3,2) model $SU(3)\times SU(2)$ without superpotential.}
{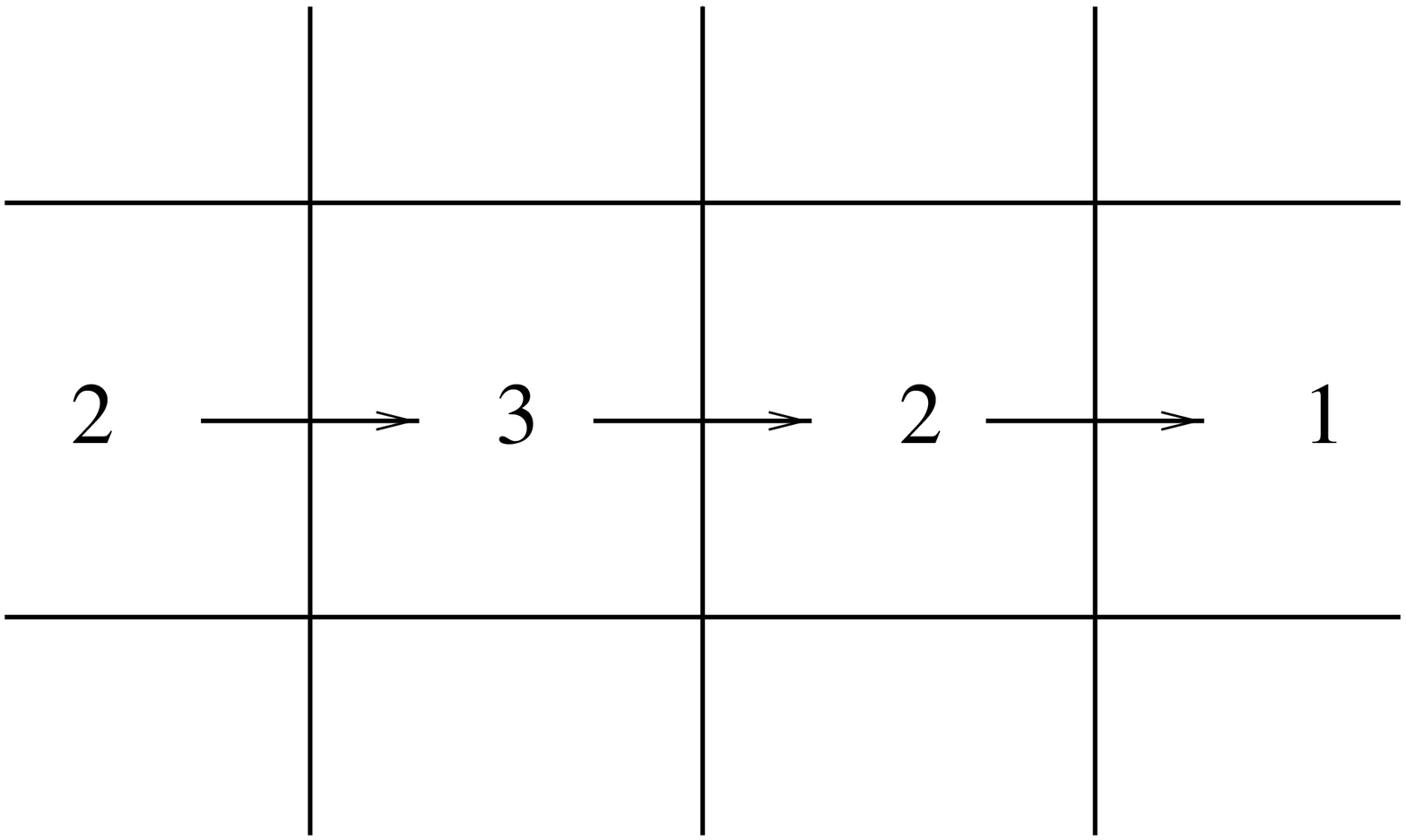}{8 truecm}
\figlabel\ttt
The particular disposition of boxes in fig. \tt\ is crucial. A theory with
the same matter content, but without superpotential, can be realized as in
fig. \ttt. The various fields are localized at different points and do not
interact. There is no superpotential and therefore we have flat directions
parametrized by the three gauge-invariant operators, $R_1R_2Q^2$, $R_1QL$ and
$R_2QL$.
They are reproduced by the possible brane motions: D4-branes can reconnect
to an infinite one, which can move in 67 or the last two NS-branes can move
with the right number of attached D4. 

The model can be immediately generalized to $SU(N)\times SU(2)$ with a similar
matter content (substitute 3 with $N$ in fig. \tt ), which is also supposed
to break supersymmetry \thintwo.
 
There is a second natural generalization of this model.
Consider general integer numbers in the four boxes in fig. \tt.
The gauge group is $SU(n_1)\times SU(n_2)$, with a global symmetry
$SU(m_1)\times SU(m_2)$. The matter content transforms in the 
\begintable
\tstrut  | $SU(n_1)$ | $SU(n_2)$ | $SU(m_1)$ | $SU(m_2)$ \crthick
$R$ | $\overline{\fund}$ | 1 | 1 | $\fund$ \cr
$Q$ | $\fund$ | $\overline{\fund}$ | 1 | 1 \cr
$L$ | 1 | $\fund$ | 1 | $\overline{\fund}$ \cr
$R'$ | $\overline{\fund}$ | 1 | $\fund$ | 1  
\endtable
Anomaly cancellation requires $m_1+m_2=n_2$ and $n_1=m_2$.
The superpotential is given by
\eqn\thtwe{W=RQL.}
There is a range of values for $n_1$ and $n_2$ for which the brane construction
lifts all
the flat directions. From the field theory point of view, baryonic flat directions may survive the introduction of the superpotential \thtwe (see \refs{\shad,\postwo}). Here the $U(1)$ factors that we assumed to be frozen may play a
role.
It would be interesting to understand this point better and to study if these theories break supersymmetry as the
(3,2) model does. 

In general, it is quite easy to construct brane configurations in which there
is no allowed motion for the branes and, therefore, there are no expected flat
directions. All these theories are, in principle, good candidates for models
that break supersymmetry.  
\subsec{Three-box models}

\fig{Various cases of three-box models.}
{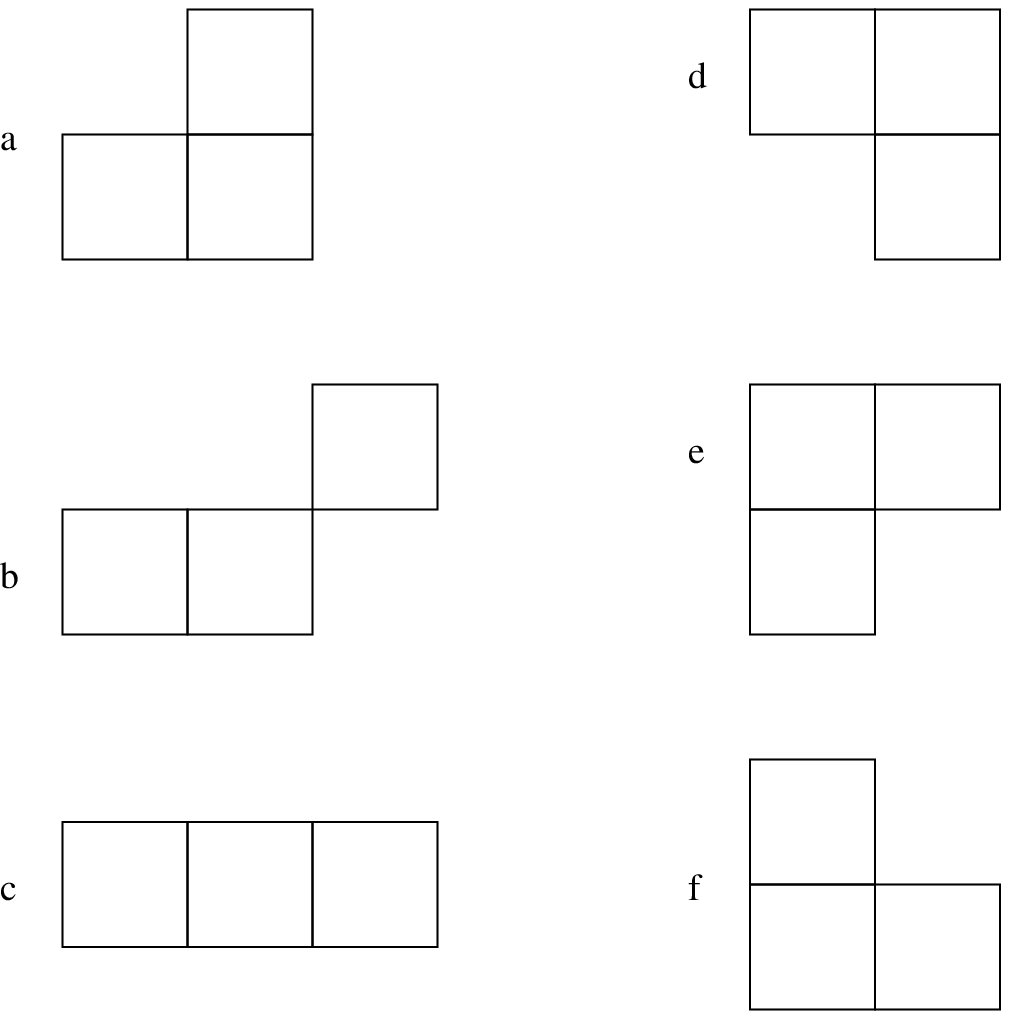}{8 truecm}
\figlabel\boxes

Just to provide more examples, we can classify all the models that can be obtained by
arranging configurations of three boxes. Figure \boxes\ presents all possible
cases.
Cases $b,d,f$ are excluded, as they are  anomalous. Cases $a,c,e$ have a gauge group
$SU(n)$ with $n_f$ flavours. Case $c$ has no superpotential. Cases $c$ and $e$
have a superpotential that gives mass to the quark fields.

To this classification, we can add an orientifold plane O7 or O7$'$ and produce
more models. The orientifold planes can be put either on top of an NS-brane or
not. 
This procedure can be repeated for a configuration of four boxes and so on.
We expect that a large class of models can be analysed systematically, using
this classification in terms of branes.

\subsec{$SU(N)$ with antisymmetric}

We are interested in an $SU(N)$ gauge theory with an antisymmetric tensor
representation $A$, $F$ fundamentals and $N+F-4$ antifundamentals. There are several
ways for realizing this theory, according to whether or not we want a
superpotential. The crucial ingredient is an NS-brane stuck at an orientifold
plane, according to the rules in section 2.

\fig{Different realizations for $SU(N)$ with an antisymmetric, $F$ fundamentals
and $N+F-4$ antifundamentals.}
{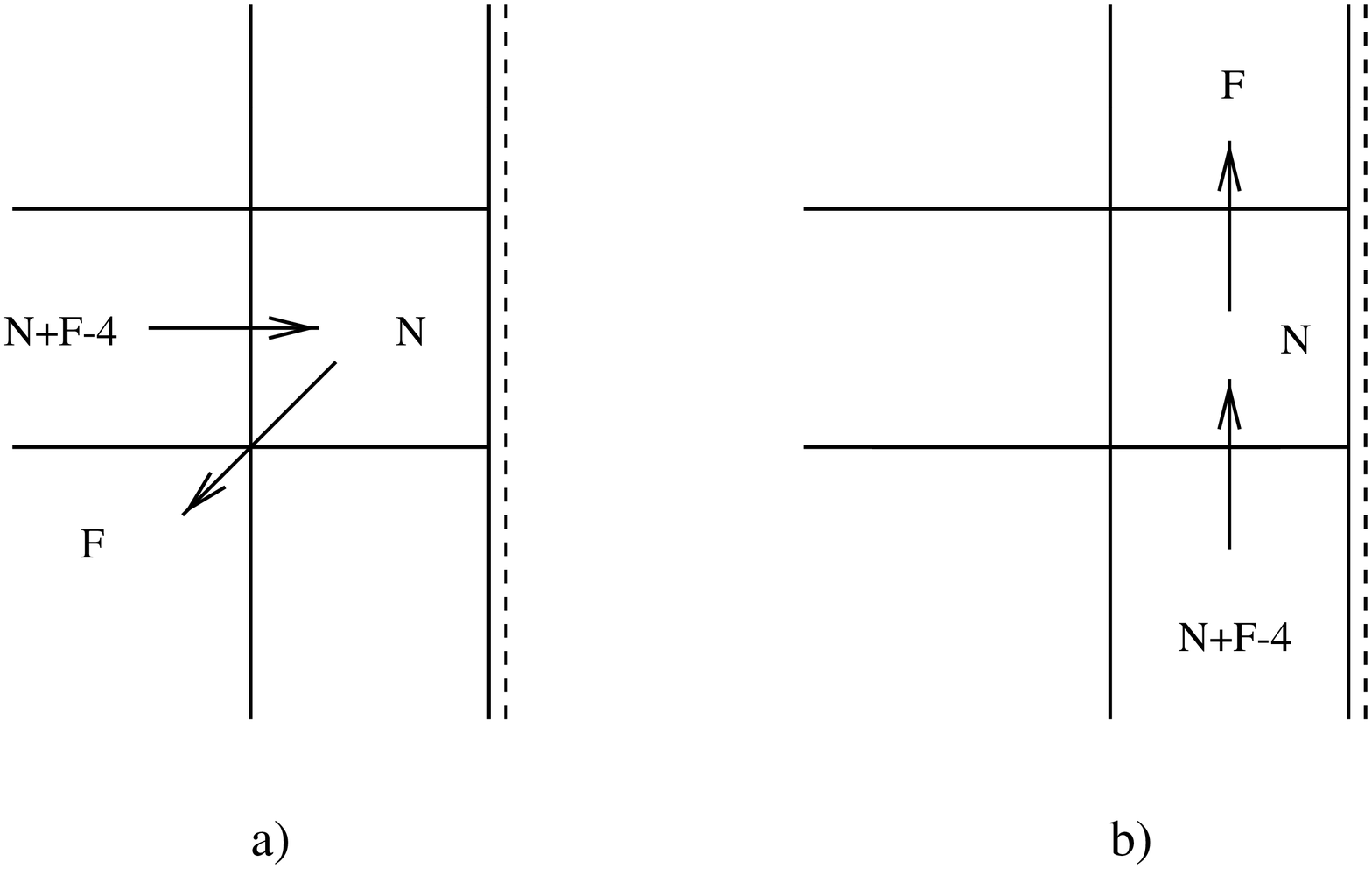}{8 truecm}
\figlabel\anti

Consider two possible realizations. The case a) in fig. \anti\ has no superpotential, while the case in figure b)
has a superpotential of the form
\eqn\antisup{W=A\bar Q\bar Q.}
This is because  the antisymmetric tensor arises from open strings localized at the orientifold plane and they can interact with the antifundamental representations only in case b) of fig. \anti.

Let us analyse the interesting case $F=0$. In case a),
we have flat directions, which correspond to reconnecting D5-branes
into infinite ones and moving them in 67. Due to the presence of the orientifold plane, we can only move away an even number of D5-branes. The theory can be higgsed to $USp(4)$ for $N$ even and $SU(5)$ for $N$ odd.  This agrees with the field theory
expectations \aff.
For $N$ odd, the model, when equipped with the superpotential \antisup, is supposed
to break supersymmetry. We can check that in case b) of fig. \anti (for $F=0$) there is only one allowed motion of branes: we can move away the NS-brane stuck at the orientifold plane. However, the resulting configuration
(see fig. 3) requires an even number of D5-branes to live at the orientifold.
This means that there exist flat directions only when $N$ is even. This is in agreement with 
the complete lifting of the flat directions in field theory and the conjectured susy breaking when $N$ is odd.

\subsec{Chiral non-chiral theories}
In this section we show that, even in our classical approximation with zero
string coupling, we can say something about the non-trivial IR properties of some
models.

It is known \berk\ that, in order to study the IR properties of a model, a
chiral theory can be expanded into a non-chiral theory with more gauge factors
but, sometimes, with simpler properties.
For example, $SU(n)$ with an antisymmetric
and $n_f=n-4$ antifundamentals can be equivalently described using the non-chiral theory $USp(n-4)\times SU(n)$ \foot{This theory can be considered chiral
in the sense that the bifundamental and the $n-4$ fundamentals are chiral, but
the number of fundamentals and antifundamentals for $SU(n)$ is the same.}
with a chiral bifundamental and $n-4$ antifundamentals for $SU(n)$.
At strong coupling the $USp$ group confines, and the mesons of the theory,
which completely saturate the anomaly, reproduce the antisymmetric
for $SU(n)$. The non-chiral model can be easily realized in terms of branes.
It indeed belongs to the infinite series of models in \vectwo.

\fig{Brane realization of $USp(n-4)\times SU(n)$ gauge group with a
bifundamental and a superpotential $W=0$.}
{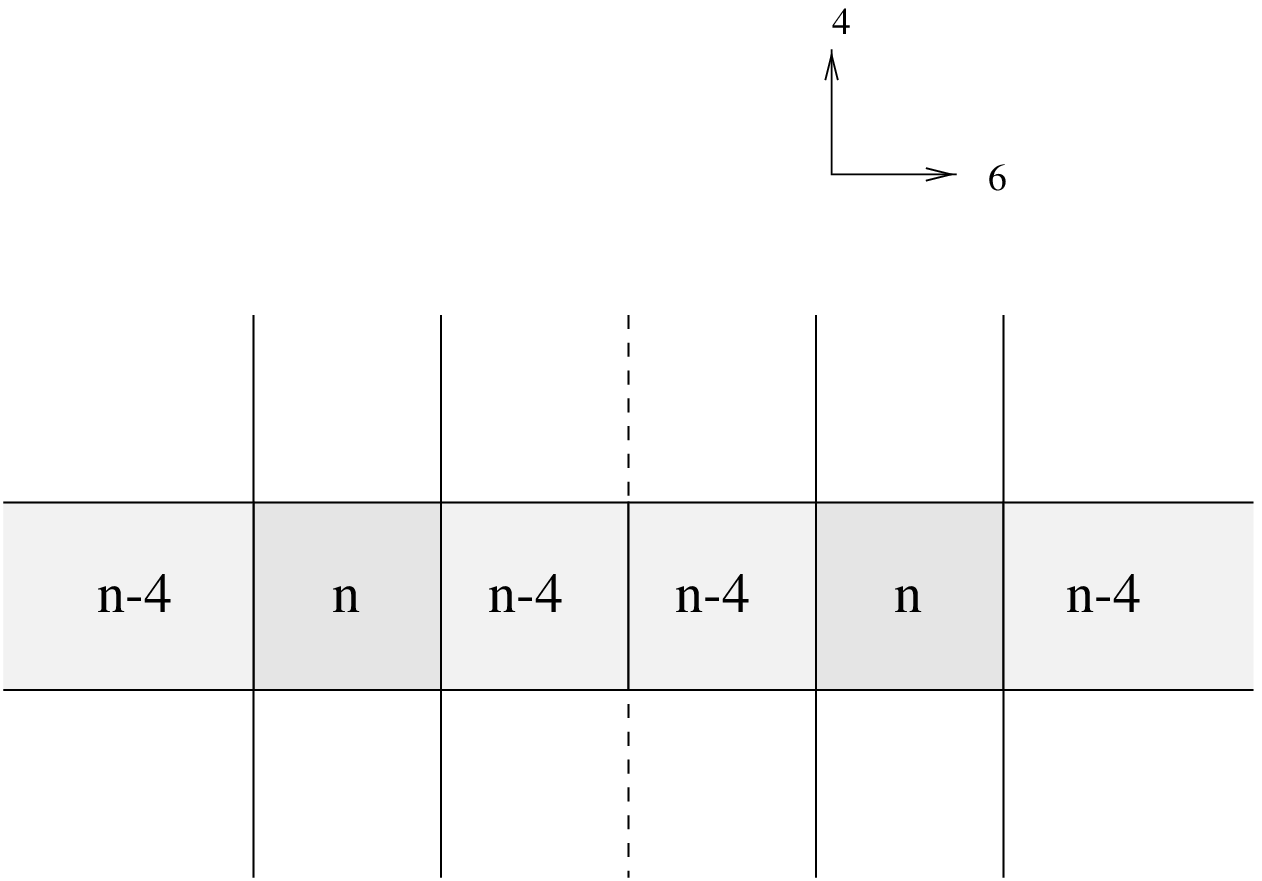}{15 truecm}
\figlabel\spsu

Consider in fact, as in fig. \spsu, the case with two NS-branes not at the
orientifold point. Stretch $n-4$
D5-branes between the first NS and its image. Using rule (b), we need $n$
D5-branes between the first and the second NS. Using rule (a), we need to
put also $n-4$ semi-infinite D5-branes. The equivalence
between the chiral and the non-chiral model can be easily shown by moving
the first NS-brane toward the orientifold.
When the NS-brane together with its image touch the orientifold, the $USp$
coupling constant flows to infinite value. One of the two NS-branes at the
orientifold is now free to move in the 789 directions. We move it to large 789
positions.
The resulting theory is therefore $SU(n)$ with an antisymmetric and
$n_f=n-4$ antifundamentals. 

A similar mechanism was discussed in \hztwo\ in the context of $(0,1)$ 
six-dimensional theories to explain the small instanton transition in
which a tensor multiplet is traded with 29 hypermultiplets. This phenomenon
was first discovered in the case of the small $E_8\times E_8$ instanton,
which has an interpretation as a M-theory five-brane which has left the
boundary, when the $E_8\times E_8$ heterotic string is interpreted as M-theory
on $S^1/Z_2$.
Also the theory of $SO(32)$ small instantons at spacetime singularities
sometimes has a Coulomb branch parametrized by tensor multiplets \intr.
If we discard the two NS$'$-branes in the models considered before and perform
T-duality in two
directions we exactly recover the theory of $SO(32)$ small instantons at
spacetime singularities as described in \hztwo. As shown there, even in this
case, the small instanton transition can be interpreted as a five-brane that has
left the boundary (in this case an O8 orientifold plane). The mechanism is
essentially the same as the one described above to demonstrate the equivalence
between chiral and non-chiral models\foot{Notice that the argument does not
use in any way the existence of the NS$'$-branes.}.

It was shown in \eva\
that the small instanton transition, when the theory is further compactified
to a four-dimensional $N=1$ model, can give rise to a transition in which the
net number of generations changes. We have explicitly seen that a non-chiral
model can be connected to a chiral one with a mechanism related by
T-duality to the six-dimensional small instanton transition.
The results in \eva\ and
those described above seem to suggest that the relation between the small
instanton transition and the physics of chirality-changing transition in four
dimensions is quite general and follows from the same universal effect.
It would be interesting to check the relation between the two approaches
and to learn more about such transitions. 

Other IR results, which we expect to be able to obtain within the classical
approximation, concern Seiberg's duality. We expect that the techniques in
\kut\ can be applied to our brane construction.

\newsec{Conclusions}
In this paper we have presented a general setup for constructing four-dimensional gauge theories, which are generically chiral.
 From the examples that we presented in the previous sections, the rules
for constructing models should now be clear.
Using these rules, we can construct quite a lot of the models present in the
literature. Many of them are supposed to break supersymmetry.
Other models, not treated in the literature, can be studied systematically.
It is also easy to construct general models without flat directions, which may
break supersymmetry.
Moreover, the brane construction arranges the various models into families
that can be treated in a unified way.
Different
dispositions of boxes gives rise to different superpotentials. In this paper we
only discussed the superpotential, which is naturally present in the brane
configuration. However, more general interactions can be obtained by introducing
more ingredients in the picture, for example rotations for some branes. 

We hope that the flexibility of this construction can be useful for constructing
and studying models when the field theory analysis gets complicated.
In this paper we limited ourselves to a classical analysis, but, even at the
classical level, the brane construction can give a help in finding the
moduli space of flat directions. The issue of supersymmetry breaking can be
addressed only when we turn on the coupling constant,
which introduces crucial differences and bending in the branes configuration.
New tools are needed to study the quantum theory. We hope to return to this
subject, clearly the most interesting one, in the future.

\centerline{\bf Acknowledgements}
We would like to thank O. Aharony, K. S. Babu, J. H. Brodie, B. Kol, L. Randall, R. Rattazzi and M. Strassler
for useful discussions.

\listrefs
\bye